\documentclass[12pt]{article}

\usepackage[fleqn]{amsmath}
\usepackage{amsfonts}
\usepackage{amssymb}
\usepackage{fullpage}
\usepackage{epsfig}

\makeatletter
\renewcommand \thesection {\@arabic\c@section}
\renewcommand\thesubsection   {\thesection.\@arabic\c@subsection}
\renewcommand\thesubsubsection{\thesubsection .\@arabic\c@subsubsection}
\renewcommand\theparagraph    {\thesubsubsection.\@arabic\c@paragraph}
\renewcommand\section{\@startsection {section}{1}{\z@}%
                                   {-3.5ex \@plus -1ex \@minus -.2ex}%
                                   {1.9ex \@plus.2ex}%
                                   {\normalfont\large\bfseries\centering}}
\renewcommand\subsection{\@startsection{subsection}{2}{\z@}%
                                     {-2ex\@plus -1ex \@minus -.2ex}%
                                     {1.2ex \@plus .2ex}%
                                     {\normalfont\normalsize\bfseries\centering}}
\renewcommand\subsubsection{\@startsection{subsubsection}{3}{\z@}%
                                     {-2ex\@plus -1ex \@minus -.2ex}%
                                     {.5ex \@plus .2ex}%
                                     {\normalfont\normalsize\em}}
\renewcommand\paragraph{\@startsection{paragraph}{4}{\z@}%
                                    {3.25ex \@plus1ex \@minus.2ex}%
                                    {-1em}%
                                    {\normalfont\normalsize\em}}
\renewcommand\subparagraph{\@startsection{subparagraph}{5}{\parindent}%
                                       {3.25ex \@plus1ex \@minus .2ex}%
                                       {-1em}%
                                      {\normalfont\normalsize\em}}
\makeatother

\DeclareMathOperator{\Artanh}{artanh}
\DeclareMathOperator{\supp}{supp}

\newcommand{\dd}{\mathrm{d}}
\newcommand{\dds}{\frac{\dd\phantom{s}}{\dd \tau}}

\newcommand{\ddt}{\frac{\dd\phantom{s}}{\dd t}}
\newcommand{\pdt}{\partial_t}

\newcommand{\tilt}{{\tilde{t}}}

\newcommand{\refeq}[1]{(\ref{#1})}

\newcommand{\Mrot}{{M}_{\text{b}}}  	
\newcommand{\rotM}{{\cal M}_{\text{b}}}	
\newcommand{\rotI}{{\cal I}_{\text{b}}}	

\newcommand{\mz}{m_{\text{b}}}
\newcommand{\Iz}{{\mathbf{I}_{\text{b}}}}

\newcommand{\labF}{{\cal F}_{\text{lab}}}
\newcommand{\LoF}{{\cal F}_{\textsc{l}}}

\newcommand{\fe}{f_{\text{e}}}
\newcommand{\fm}{f_{\text{m}}}

\newcommand{\vect}[1] {\boldsymbol{{ #1}} }

\newcommand{\qv}[1]{{\textbf{\textsl{#1}}}}

\newcommand{\tenseur}[1]{{\textbf{\textsf{#1}}}}

\newcommand{\Rset}{\mathbb{R}}

\newcommand{\AQ}{\qv{A}}                

\newcommand{\FQ}{\tenseur{F}}           
\newcommand{\gQ}{\tenseur{g}}           
\newcommand{\SQb}{{\tenseur{S}_{\text{b}}}} 
\newcommand{\tQT}{\tenseur{t}}          

\newcommand{\aQ}{\qv{a}}                
\newcommand{\eQ}{\qv{e}}                
\newcommand{\fQ}{\qv{f}}        	
\newcommand{\JQ}{\qv{J}}                
\newcommand{\pQ}{\qv{p}}                
\newcommand{\uQ}{\qv{u}}                
\newcommand{\xQ}{\qv{x}}                
\newcommand{\yQ}{\qv{y}}                
\newcommand{\zQ}{\qv{q}}                

\newcommand{\aV}{\vect{a}}              
\newcommand{\jV}{{\vect{j}}}		
\newcommand{\sV}{{\vect{s}}}            
\newcommand{\sVb}{{\vect{s}}_{\text{b}}} 
\newcommand{\sVe}{{\vect{s}}_{\text{f}}}
\newcommand{\tV}{\vect{t}}              
\newcommand{\uV}{{\vect{u}}}            
\newcommand{\vV}{{\vect{v}}}            
\newcommand{\xV}{\vect{x}}              
\newcommand{\yV}{\vect{y}}              

\newcommand{\sbstar}{{s^\sharp_{\text{b}}}}

\newcommand{\AV}{\pmb{{\cal A}}}
\newcommand{\BV}{\pmb{{\cal B}}}
\newcommand{\EV}{\pmb{{\cal E}}}
\newcommand{\FV}{\pmb{{\cal F}}}
\newcommand{\GV}{\pmb{{\cal G}}}
\newcommand{\LV}{\pmb{{\cal L}}}
\newcommand{\PV}{\pmb{{\cal P}}}

\newcommand{\WV}{\pmb{{\cal W}}}

\newcommand{\nab}{\vect{\nabla}}
\newcommand{\nabQ}{\nab_{\text{g}}}
\newcommand{\lapQ}{{\Delta}_{\text{g}}}
\newcommand{\wop}{{\square}}

\newcommand{\EulerQ}{\mbox{$\vect{\Omega}_{\text{\textsc{e}}}$}}

\newcommand{\eulerV}{\mbox{$\vect\omega$}}
\newcommand{\euler}{\mbox{$\omega$}} 
\newcommand{\OmegaQ}{\vect{\Omega}}
\newcommand{\omegaQ}{\mbox{$\qv{w}$}}
\newcommand{\omegaV}{\vect{\omega}}

\newcommand{\Mmink}{\tenseur{M}} 		
\newcommand{\Mnodv}{\tenseur{M}_{\text{\textsc{n}}}} 
\newcommand{\RC}{R_{\textsc{c}}}        

\newcommand{\qdot}[1]{{\stackrel{\,\circ}{{#1}}}}

\newcommand{\abs}[1]{\left| #1 \right|}
\newcommand{\norm}[1]{\left\| #1 \right\| }
\newcommand{\norml}[1]{\left\| #1 \right\|_{1,\lambda}}

\newcommand{\defeg}{\stackrel{\textrm{\tiny def}}{=}}

\renewcommand{\leq}{\leqslant}
\renewcommand{\geq}{\geqslant}

\reversemarginpar

\numberwithin{equation}{section}

\begin{document}



\title{\uppercase{
	Scattering\, and\, Radiation\, Damping\, in \\
	Gyroscopic Lorentz Electrodynamics}}

\author{
\textbf{Walter APPEL}$^{*,\dagger}$ and \textbf{Michael K.-H. KIESSLING}\\
                Department of Mathematics, Rutgers University\\
                110 Frelinghuysen Rd., PISCATAWAY, NJ 08854, USA\\ 
        $^*$\emph{On leave from}: Laboratoire de Physique\\
                Unit\'e de recherche 1325 associ\'ee au CNRS\\
                \'Ecole normale sup\'erieure de Lyon\\
                46 all\'ee d'Italie, 69\,364 LYON Cedex 07 France\\ 
	$^\dagger$\emph{Present address}: Lyc\'ee Henri Poincar\'e \\
	54\,000 Nancy, France\\ \\
\textrm{Version of Jan. 04, 2002. Printout typeset with \LaTeX\ on}}
\maketitle
\thispagestyle{empty}

\begin{abstract}
{\noindent 
	The nonlinear system of equations of relativistic 
Lorentz electrodynamics (LED) is studied in a ``gyroscopic setup'' 
in which the Lorentz electron is assumed to remain at rest, 
leaving the electromagnetic fields and the particle spin as the only 
dynamical degrees of freedom. 
	The global existence and uniqueness of this gyroscopic
spin-plus-field dynamics in unbounded space is proven.
	It is further shown that for rotation-reflection symmetric 
initial data any gyroscopic solution also satisfies the world-line 
equations consistent with a non-moving Lorentz electron, thus furnishing 
a proper solution of the complete set of equations of LED.
	Rotation-reflection symmetric scattering is shown to occur for 
sufficiently small ratio of electrostatic to (positive) bare rest mass, with 
deviations from the stationary spin state dying out exponentially fast 
through radiation damping. 
	The previously proven result that the renormalized spinning 
Lorentz electron evolves like a soliton in scattering processes
combined with the present results that scattering does occur
establish the solitonic character of the renormalized Lorentz electron.
$$
\mathrm{To\ appear\ in:}\ \mathbf{Letters\ in\ Mathematical\ Physics}
$$
}
\end{abstract}

\vfill
\hrule
\smallskip
\noindent
\textrm{\copyright 2001. The authors. Reproduction for non-commercial
purposes, by any means, is permitted.}

\newpage




                \section{Introduction}
	In recent years the century-old Lorentz program of 
electrodynamics~\cite{rohrlichBOOK} has attracted the
attention of mathematical physicists.
	Most of the rigorous results established so far belong to the
semi-relativistic Abraham model and are surveyed in~\cite{spohn}.
	Recently~\cite{appelkiesslingAOP} the authors presented the 
first properly renormalized approach to truly relativistic 
Lorentz electrodynamics (LED), picking up on the pioneering 
work~\cite{nodvik}. 
	While the  dynamical initial value problem for the 
model in~\cite{nodvik} is seriously singular, our Lorentz-covariant 
LED displays most of the features considered crucial for a realistic, 
consistent classical electrodynamics, namely:

	$\bullet$
the Cauchy problem for the evolution of the physical state in 
``massive'' LED with strictly 

positive bare rest mass and bare moment of inertia is regular;

	$\bullet$
the pre- and post-scattering values of the renormalized electron rest
mass and electron spin

magnitude are identical, i.e. the Lorentz electron evolves 
as soliton in scattering processes;

	$\bullet$
there exists a simple curve in the charge -- bare rest mass --
radius -- gyration frequency ---

parameter space of the stationary bare particle on which
the stationary renormalized particle 

data charge, magnetic moment, and  mass  match the empirical 
electron data without involving 

superluminal gyration speeds.

\noindent
	In~\cite{appelkiesslingAOP} we also studied  LED's
renormalization flow to vanishing bare rest mass with empirically matched 
data when the positive bare mass and charge are distributed on the surface of 
a sphere.
	The renormalized ``purely electromagnetic'' LED which emerges in 
the limit has the following additional characteristics:

	$\bullet$
the renormalized purely electromagnetic LED constitutes a classical field 
theory equipped 

with an ultraviolet cutoff at about the  physical electron's Compton length;

	$\bullet$
in the limit of vanishing bare rest mass the equatorial gyration speed 
reaches the speed of 

light and the bare gyrational mass converges to a ``photonic'' mass;

	$\bullet$
in the same limit, the renormalized spin magnitude converges to 
$3\hbar/2$, up to corrections of 

order $\alpha$ (Sommerfeld's fine structure constant).

	In this paper we supply several rigorous results regarding 	
scattering in LED conditioned on a straight particle world-line.
	For a straight particle world-line the set of Maxwell--Lorentz 
plus gyrational equations decouples from the world-line equations, which 
in turn become constraint equations that impose on the admissible set of 
initial conditions for the dynamical equations for the
spin and the electromagnetic fields.
	We prove that all physically reasonable Cauchy data for the fields 
and the spin launch unique global forward and backward evolutions 
of Maxwell--Lorentz plus gyrational equations.
	For rotation-reflection symmetric  data it is shown that these
gyroscopic solutions satisfy the world-line equations for a non-moving 
Lorentz electron, too, thus furnishing solutions of the complete set of 
equations of LED. 
	Rotation-reflection symmetric scattering is shown to occur if 
the ratio of electrostatic to bare rest mass is smaller than $\approx 1$.
	The previously proven result that the renormalized spinning 
Lorentz electron evolves like a soliton in scattering
processes~\cite{appelkiesslingAOP} combined with the present results 
that scattering does occur establish the solitonic character of the 
renormalized Lorentz electron.
	It is further shown that the rotation-reflection symmetric deviations 
from the soliton state die out exponentially fast through radiation damping.
	The results proven here are somewhat stronger and cover 
more general mass and charge densities than announced 
in~\cite{appelkiesslingAOP}.
 
    \section{Notation}
    \label{sec:PREP}
	We use the notation of~\cite{appelkiesslingAOP},
which largely follows the conventions of~\cite{misnerthornewheeler}.  
	Abstract Minkowski space is identified with $\Rset^{1,3}$, 
equipped with a Lorentzian metric of signature $+2$.  
        Thus, any orthonormal basis $\{\eQ_0,\eQ_1,\eQ_2,\eQ_3\}$
satisfies the elementary inner product rules
$ \eQ_0\cdot \eQ_0 = -1$, $\eQ_\mu\cdot \eQ_\mu =1$ for
$\mu > 0 $, and $\eQ_\mu\cdot \eQ_\nu  =  0$ for $\mu\neq \nu$. 
	A constant basis defines a \emph{Lorentz frame}, denoted $\LoF$.
	We use $\qv{x}$, $\qv{y}$, etc. to denote four-vectors 
representing events in \emph{spacetime}. 
	With respect to $\LoF$, we decompose $\qv{x}$ 
into time-plus-space components thus, $\qv{x} = (ct, \vect{x})$, 
where $\vect{x}= (x^1,x^2,x^3)$ is a ``point in space,'' 
and $t=x^0/c$ an ``instant of time,'' where $c$ is the speed of light
\emph{in vacuo}.
	Henceforth we shall use units in which $c=1$.
        We call $\qv{v}$ spacelike, lightlike, or timelike according 
as $\qv{v}\cdot\qv{v} >0$, $\qv{v}\cdot\qv{v} =0$, or $\qv{v}\cdot\qv{v} <0$, 
respectively.
        We define $\norm{\qv{v}}$ as the principal value of 
$(\qv{v}\cdot\qv{v})^{1/2}$.  
	The tensor product $\eQ_\mu\otimes \eQ_\nu$
is defined by its inner-product action on four-vectors thus, 
$(\eQ_\mu\otimes \eQ_\nu)\cdot \qv{c}\defeg  \eQ_\mu  (\eQ_\nu\cdot \qv{c})$
and $\qv{c}\cdot(\eQ_\mu\otimes \eQ_\nu)\defeg (\eQ_\mu \cdot \qv{c})\eQ_\nu$.
	In general a rank-two tensor reads
$\tenseur{T}  =  T^{\mu\nu}\eQ_\mu \otimes \eQ_\nu$, and 
if $T^{\mu\nu}= \pm T^{\nu\mu}$ it is 
\emph{symmetric} ($+$ sign), respectively \emph{anti-symmetric} ($-$ sign).
	The \emph{metric tensor} $\gQ = g^{\mu\nu}\eQ_\mu \otimes \eQ_\nu$, 
with $g^{\mu\nu} = \eQ_\mu\cdot \eQ_\nu$, is clearly symmetric and
has the same components $g^{\mu\nu}$ in all Lorentz frames.
	Notice that $\gQ$ acts as identity on 
four-vectors, i.e. $\gQ\cdot\qv{v} = \qv{v}$.
	A particular class of anti-symmetric tensors  is given
by the exterior product between two four-vectors,
$\qv{a}\wedge\qv{b}\defeg \qv{a}\otimes \qv{b} -  \qv{b}\otimes \qv{a}$.
	Finally,  $\left[\tenseur{A},\tenseur{B}\right]_\pm\defeg
        	\tenseur{A}\cdot\tenseur{B} \pm \tenseur{B}\cdot\tenseur{A}$
is the \emph{(anti-)commutator} of any two tensors of rank two
$\tenseur{A}$ and $\tenseur{B}$.

        For a differentiable function $f(\qv{x})$ we denote by $\nabQ{f}$ 
its four-gradient w.r.t. $\tenseur{g}$.
	In time-plus-space decomposition,
$\nabQ f(\xQ)= \left(-\partial_{_{x^0}}f,\nab f\right)$,
where $\nab$ is the usual three-gradient. 
	The four-curl of a differentiable four-vector function 
is defined in analogy with the conventional curl as the 
anti-symmetric four tensor function
\begin{equation}
        \nabQ \wedge \qv{A}(\xQ)\
=       
	\varepsilon^{\mu\nu\lambda\eta}
	\eQ_\mu \otimes \eQ_\nu (\eQ_\lambda\cdot\nabQ)( \eQ_\eta \cdot\qv{A})
\end{equation}
where the $\varepsilon^{\mu\nu\lambda\eta}$ are the  entries of the 
rank-four Levi-Civita tensor.
        The four-Laplacian with respect to $\tenseur{g}$
is just the (negative) d'Alembertian, or
wave operator, i.e. $ \lapQ\defeg\nabQ \cdot \nabQ =  -\wop$.

   \section{Covariant massive  LED with a straight particle world-line}

	In this section we present the manifestly covariant
equations of massive LED~\cite{appelkiesslingAOP} for the 
special case that the particle's world-line is straight.

	\subsection{Kinematical pre-requisites} 
	We recall that the particle's \emph{world-line} is a map
$\tau\mapsto\xQ =\zQ(\tau)$, where
$\dd \tau\ = \sqrt{- \dd \xQ\cdot\dd \xQ}$, with $\dd\xQ$ taken 
along the world-line, is the invariant \emph{proper-time} element. 
	The map $\tau\mapsto\uQ =\qdot{\zQ}(\tau)$, where
$\qdot{\zQ}$ is the particle's four-velocity, is the 
\emph{world-hodograph}.
	The \emph{world-gyrograph} of the particle 
is an anti-symmetric tensor-valued map $\tau\mapsto\EulerQ(\tau)$ of
space-space type with respect to $\uQ$ (i.e. $\EulerQ\cdot \qv{u}=\qv{0}$) 
which describes the angular velocity of the inert gyrational motions of the 
Lorentz particle that may occur in excess of the inertia-free 
{Thomas precession}.
	Thus, $\EulerQ \defeg \OmegaQ- \qdot{\uQ}\, \wedge\,\uQ$, 
where $\OmegaQ$ is the angular velocity tensor of 
the particle's co-rotating body frame, while $\qdot{\uQ}\, \wedge\,\uQ$ 
is the familiar angular velocity tensor of Fermi--Walker
transport~\cite{misnerthornewheeler}.

	For a straight world-line $\zQ(\tau) = \uQ_0\tau + \zQ_0$
the particle's four-velocity is a constant four-vector,
$\uQ(\tau)=\uQ_0$ for all $\tau$. 
	A constant four-velocity in turn implies that $\EulerQ = \OmegaQ$.

	\subsection{Field equations}
	The electromagnetic Maxwell--Lorentz fields are gathered
into the anti-symmetric rank-two Faraday tensor field
$\xQ\mapsto\FQ(\xQ)$, which satisfies the manifestly 
covariant Maxwell--Lorentz equations
\begin{equation}
        \nabQ\, \cdot\, {^\star\FQ} 
=
	\qv{0}
\, ,\label{eq:homMLeq}
\end{equation}
\begin{equation}
        \nabQ\cdot \FQ 
=
        4\pi \JQ\, ,
\label{eq:inhMLeq}
\end{equation}
where ${^\star\FQ}$ is the (left) Hodge dual of $\FQ$ and $\JQ$ is 
the charge-current density four-vector field, given by
Nodvik's~\cite{nodvik} manifestly covariant expression
\begin{equation}
        \JQ(\xQ)
=
        \int_{-\infty}^{+\infty}
		(\uQ_0 -\OmegaQ(\tau)\cdot \qv{x})
        	\;{\fe}\big(\norm{\qv{x}-\zQ(\tau)}\big)
	        \,\delta\big(\uQ_0\cdot(\qv{x}-\zQ_0)+\tau\big)
	\,\dd \tau
\, ,\label{eq:Jnodvik}
\end{equation}
where $\fe: [0,R]\to\Rset^-$ is the $SO(3)$ invariant charge
``density'' of the Lorentz particle, and $0<R<\infty$ its radius.
	For a Lorentz electron, $\int_{\Rset^3}\fe(|\xV|)\dd^3x = -e$, 
where $e>0$ the elementary charge.
	Conditioned on the world-line 
$\tau\mapsto \zQ(\tau)= \uQ_0\tau + \zQ_0$ and gyrograph 
$\tau\mapsto\EulerQ(\tau) = \OmegaQ(\tau)$ being given, the
Maxwell--Lorentz equations are linear equations for $\FQ$. 
	\subsection{World-gyrograph equations}
	The  equations for the gyrograph are 
\begin{equation}
	\dds {\SQb}  
= 
	\tQT
\, ,  \label{eq:ggEQ}
\end{equation}
where
\begin{equation}
\!\!\!\! 
	\SQb(\tau)
=
	\int_{\Rset^{1,3}}\!\!
	\big(\yQ-\uQ_0\tau\big)\wedge 
	 \frac{-\EulerQ(\tau)\cdot \yQ}
        {\sqrt{\displaystyle 1-\norm{\EulerQ(\tau)\cdot\yQ}^2}}
        \fm\big(\norm{\yQ-\uQ_0\tau}\big)\,\delta\big(\uQ_0\cdot\yQ+\tau\big)\,
        \dd^4y
\label{eq:barespinTENSOR}
\end{equation}
is the anti-symmetric tensor of \emph{bare Minkowski spin (about 
$\zQ(\tau)=\uQ_0\tau +\zQ_0$)} associated with the gyrational motion of
the $SO(3)$ invariant bare rest mass ``density''
$\fm: [0,R]\to\Rset^+$ of the particle, while
\begin{equation}
\!\!\!\!\!\!\!\!
	\tQT(\tau) 
=\!
	\int_{\Rset^{1,3}}\!\!
		\big(\yQ-\uQ_0\tau\big)\!\wedge\! 
	\bigl(\FQ(\yQ)\cdot (\uQ_0 -\OmegaQ(\tau)\cdot \qv{y})\bigr)^\perp
	\fe\big(\norm{\yQ-\uQ_0\tau}\big)\,
	\delta\big(\uQ_0\cdot\yQ +\tau\big)
	\dd^4y
\label{eq:MinkALtorque}
\end{equation}
is the Abraham--Lorentz type \emph{Minkowski torque}, with 
$\aQ^\perp\defeg \bigl(\gQ + \uQ_0\otimes\uQ_0\bigr)\cdot\aQ$.
	\subsection{World-line equations}
	The world-line equations are 
\begin{equation}
        \frac{\dd }{\dd \tau}\pQ
=
	\fQ
\, ,\label{eq:wlEQ}
\end{equation}
where
\begin{equation}
        \pQ(\tau) 
=
        \Mmink(\tau) \cdot \uQ_0
\, \label{eq:MinkEnMom}
\end{equation}
is the \emph{Minkowski momentum} four-vector of the particle, with 
$\Mmink = \Mnodv + \Mrot \gQ$ its symmetric \emph{Minkowski tensor mass}, 
where 
\begin{equation}
\!\!\!\!\!\!\!\!\!\!\!\!\!\!
\Mnodv (\tau)\!  =\! - \int_{\Rset^{1,3}}
        	\left[(\yQ-\uQ_0\tau)\otimes(\yQ-\uQ_0\tau), 
			\left[ \FQ(\yQ),\EulerQ(\tau) \right]_+ 
		\right]_+ \fe\big(\norm{\yQ-\uQ_0\tau}\big)\,
        	\delta(\uQ_0 \cdot \yQ+\tau)\dd^4y
\end{equation}
is the \emph{Nodvik tensor mass}~\cite{appelkiesslingAOP},
extracted from the Minkowski momentum four-vector associated with
electromagnetic spin-orbit coupling given in~\cite{nodvik}, and where 
\begin{equation}
	\Mrot(\tau)
 =
	\int_{\Rset^{1,3}} 
	\Big(\displaystyle{1-\norm{\EulerQ\cdot\yQ}^2}\Big)^{-\frac{1}{2}}
   \fm\big(\norm{\qv{y}-\uQ_0\tau}\big)\,\delta\big(\uQ_0\cdot\yQ+\tau\big)
	\,\dd^4y\,
\label{eq:gyroMASS}
\end{equation}
is the \emph{gyrational bare mass}~\cite{appelkiesslingAOP}. 
	Finally, 
\begin{equation}
	\fQ(\tau)
= 
      \int_{\Rset^{1,3}} \FQ(\yQ)\cdot
(\uQ_0 -\EulerQ(\tau)\cdot \qv{y})
        \fe(\norm{\yQ- \uQ_0\tau})
	\,\delta\big(\uQ_0\cdot\yQ +\tau\big)\,\dd^4y
\label{eq:MinkALforce}
\end{equation}
is the Abraham--Lorentz type \emph{Minkowski force}~\cite{nodvik}.
        \section{The Cauchy problem for the state in LED}
	We now choose a convenient Lorentz frame, called the 
``laboratory frame'' $\labF$, in which the space-plus-time 
decomposition of our manifestly covariant equations takes a simple form.
	In particular, since we consider only evolutions for which
$\uQ(\tau) = \uQ_0$ for all $\tau$, we can work with the standard
foliation of space-time in our frame $\labF$. 
	The standard foliation of $\labF$ consists of the 
level sets $T_{\labF}(\xQ) = t$ of the function 
$T_{\labF}(\xQ)\defeg -\eQ_0\cdot\xQ$, which has a constant timelike 
four-gradient $\nabQ T_{\labF}(\xQ) = -\eQ_0$. 
	The space-plus-time decomposition of events in $\labF$
written as $(t,\vect{x})$, is understood w.r.t. this standard
foliation.

	By a boost we can achieve that the timelike unit 
vector $\eQ_0= (1,0,0,0)$ of $\labF$ coincides with the four-velocity 
of the particle, i.e. $\uQ_0 =\eQ_0$. 
	By at most a spacetime translation we can furthermore assume 
that  $\zQ(0) = \qv{0}$ in $\labF$, so that the particle's 
space position is at the origin of the space hypersurface of $\labF$, 
and that laboratory time $t$ and particle proper-time $\tau$ coincide.
	Accordingly, from now on we will write $t$ in place of $\tau$.
	The world-line as seen in $\labF$ is now simply given by
$\zQ(t) = (t,\vect{0})$. 
	As for the gyrograph, since $\EulerQ=\OmegaQ$, we will 
henceforth simply omit the subscript $_E$.
	In $\labF$ 
we clearly have $\OmegaQ(t)\cdot \eQ_0 =\qv{0}$ for all $t$, so 
that $\OmegaQ$ is dual to a spacelike four-vector $\omegaQ(t)$ 
which satisfies $\OmegaQ(t)\cdot \omegaQ(t) =\qv{0}$ 
and $\omegaQ(t)\cdot\eQ_0 = \qv{0}$ for all $t$.
        Hence, in our $\labF$ we have $\omegaQ =(0,\omegaV)$, 
where $\omegaV(t)$ is the usual angular velocity three-vector, 
directed along the instantaneous (i.e., at time $t$) axis of body 
gyration in the space hypersurface of~$\labF$.
	Finally, the field tensor $\FQ(\xQ)$ at $\xQ$
is decomposed as usual into its electric and magnetic 
Maxwell--Lorentz components w.r.t. the standard foliation 
of $\labF$, here conveniently grouped together as a 
complex electromagnetic three-vector field,
\begin{equation}
	\GV(\vect{x},t)
\defeg 
	\EV(\vect{x},t) + i\BV(\vect{x},t)
\, , \label{eq:EBfoli}
\end{equation}
whose real and imaginary part are, respectively, the electric
(i.e. time-space) and magnetic (i.e. space-space) components 
of the field tensor $\FQ$ in $\labF$.
	Since by hypothesis $\zQ(t) = (t,\vect{0})$ for all
$t$, the state at time $t$ in LED is uniquely characterized by
specifying $\eulerV(t)$ and $\GV(\, .\, ,t)$.
        \subsection{Evolution equations}
	The covariant equations now decompose
into a system of first-order evolution equations for the state variables
of LED, plus a set of constraint equations.
	We begin with the evolution equations.

        \subsubsection{Field equations}
	Beginning with the covariant field equations, we note
the space-plus-time decomposition of the current density four-vector
as $\JQ(\xQ) = (1,\eulerV(t)\times \vect{x}){\fe}\big(|\vect{x}|\big)$. 
	The space components of the covariant field equations combine 
into the Maxwell--Lorentz evolution equations for $\GV$,
\begin{equation}
	\partial_t\GV (\xV,t)
= 
	-i\nab\times \GV(\xV,t) - 4\pi
\,\eulerV(t)\times \vect{x}\,{\fe}\big(|\vect{x}|\big)
\, , \label{eq:MLevolvEQfoli}
\end{equation}
where $\partial_t$ means first-order partial derivative w.r.t. Lorentz 
time and $\nab\times$ is the standard curl operator.

        \subsubsection{Spin equations}

	Turning next to the gyrational equations, we recall that
$\EulerQ$ is dual to the space vector $\eulerV$.
	In the same vein, the space projector $\gQ + \uQ_0\otimes\uQ_0$ 
under the integral in \refeq{eq:MinkALtorque} guarantees the space-space 
character of $\SQb$ w.r.t. $\uQ_0$, i.e. $\SQb\cdot\uQ_0 = \qv{0}$ 
for all $\tau$, so that the bare spin Minkowski tensor 
\refeq{eq:barespinTENSOR} is dual to the space vector of bare spin,
\begin{equation}
	\sVb(t)
=  
        \int_{\Rset^3} 
        \frac{\xV\times(\eulerV(t)\times\xV)}
        {\sqrt{1-|\eulerV(t)\times\xV|^2}}{\fm}(|\xV|)\,\dd^3x 
\, ,\label{eq:barespinVECT}
\end{equation}
and the Minkowski torque \refeq{eq:MinkALtorque} is dual to the 
torque space vector
\begin{equation}
	\tV(t)
=  
        \int_{\Rset^3} 
	\xV\times \Big(\EV(\xV,t) + (\eulerV(t)\times\xV)\times\BV(\xV,t)\Big)
	{\fe}(|\xV|)\,\dd^3x 
\, .\label{eq:torqueVECT}
\end{equation}
	Equation \refeq{eq:ggEQ} together with \refeq{eq:MinkALtorque}
is therefore dual to the evolution equation 
\begin{equation}
	\ddt {\sVb}  
=
	\tV
\label{eq:dualggEQ}
\, ,
\end{equation}
for $\eulerV(t)$. 

        \subsection{Constraint equations}

        \subsubsection{Divergence equations}
	The time components of the covariant field equations 
combine into the Maxwell--Lorentz divergence equation
\begin{equation}
	\nab\cdot\GV (\xV,t)
= 
	4\pi\,{\fe}\big(|\vect{x}|\big)
\, . \label{eq:MLdivEQfoli}
\end{equation}
	Notice that \refeq{eq:MLdivEQfoli} is  merely a
constraint on the set of initial data, for the 
(three-) divergence of \refeq{eq:MLevolvEQfoli} implies that
a solution $\GV(\xV,t)$ of \refeq{eq:MLevolvEQfoli} for 
given $\eulerV(t)\times \xV\, \fe(|\xV|)$ automatically 
satisfies \refeq{eq:MLdivEQfoli} for all $t>0$ if the initial data
$\GV_0\defeg \EV_0 + i\BV_0$ satisfy the constraint \refeq{eq:MLdivEQfoli} 
at time $t=0$, i.e.
if $\nab\cdot\GV_0 (\xV)=4\pi\,{\fe}\big(|\vect{x}|\big)$. 

        \subsubsection{World-line equations}
	The four-momentum  $\pQ$ has the space-plus-time 
decomposition $\pQ = (\Mrot,\mathbf{N}_{\text{e}}\cdot\eulerV)$, where 
\begin{equation}
	\Mrot(t)
=  
        \int_{\Rset^3} 
        \frac{1}{\sqrt{1-|\eulerV(t)\times\xV|^2}}
                {\fm}(|\xV|)\,\dd^3x 
\, \label{eq:gyroMASSeval}
\end{equation}
is the bare gyrational mass at time $t$, and where 
\begin{equation}
	 \mathbf{N}_{\text{e}}(t)
 =
	\int_{\Rset^3}
	\xV\otimes\Big(\xV\times\EV(\xV,t)\Big)\fe(|\xV|)\dd^3x
\,  \label{eq:spinorbitCOUPLE}
\end{equation}
is a spin-orbit coupling tensor.
	Furthermore, the Abraham--Lorentz type Minkowski force 
now has the space-plus-time decomposition $\fQ = (P,\vect{f})$,
where
\begin{equation}
	P(t)
=	
	\eulerV(t)\cdot\int_{\Rset^3}
	\Big(\xV\times\EV(\xV,t)\Big)\fe(|\xV|)\dd^3x
\,  \label{eq:LEISTUNG}
\end{equation}
is the power delivered by the field to the particle, and where
\begin{equation}
	\vect{f}(t)
= 
      \int_{{\Rset}^{3}} 
	\Big(\EV(\xV,t) + \big(\eulerV(t)\times \xV\big)\times\BV(\xV,t)\Big)
        \fe(|\xV|) \,\dd^3x
\label{eq:ALforce}
\end{equation}
is the Abraham--Lorentz force on the particle.
	The space-plus-time decomposition of the 
world-line equation then becomes
\begin{equation}
	\ddt \Mrot
=
	P
\label{eq:powerEQ}
\, 
\end{equation}
and
\begin{equation}
	\ddt \left(\mathbf{N}_{\text{e}}\cdot\eulerV\right)
=
	\vect{f}
\label{eq:NewtonEQ}
\, .
\end{equation}
	Despite their appearance, equations \refeq{eq:powerEQ}
and \refeq{eq:NewtonEQ} are \emph{not} evolution equations for the 
world-line; instead, they have to be satisfied by the active state 
variables $\eulerV(t)$ and $\GV(\,.\,,t)$ to ensure consistency with 
the constraint that the world-line is given by $\zQ(t) = \eQ_0t$ in $\labF$.
	However, we shall show that \refeq{eq:powerEQ} is automatically 
satisfied for all time by any solution of the evolution equations for 
spin and fields that obeys the divergence equations initially.
	This leaves \refeq{eq:NewtonEQ} as the only true constraint
equation coming from the world-line equation. 
	While we will show that certain symmetric initial
conditions launch a dynamics consistent with \refeq{eq:NewtonEQ}, 
it seems difficult to precisely characterize the complete set
of initial conditions that will launch such a consistent dynamics.

        \subsection{Cauchy data}
	The field evolution equation \refeq{eq:MLevolvEQfoli} are 
supplemented by initial data consistent with the constraint equations 
\refeq{eq:MLdivEQfoli}
and satisfying the asymptotic condition that $\GV(\xV,t)\to\vect{0}$ 
as $|\xV| \to \infty$, the real part as 
$\EV(\xV,t) \sim - e \xV/|\xV|^3 +o(|\xV|^{-2})$,
the imaginary part satisfying $|\BV| = O\big(|\xV|^{-3}\big)$.

	Equation \refeq{eq:dualggEQ} is to be supplemented
by initial data $\eulerV(0)=\eulerV_0$ satisfying the
requirement of \emph{strict subluminality}, $|\eulerV_0|R<1$,
or \emph{subluminality}, $|\eulerV_0|R\leq 1$, depending
on the choice of $\fm$.

	Viewed from a dynamical systems perspective, Cauchy data 
may be prescribed in any consistent manner, and for our existence 
and uniqueness result of a strong solution in some weighted $L^1$ 
norm we only need that the cumulative time integral of the wave 
fields over the support of the particle stays bounded.
	However, the scope of LED as a theory, in the classical 
limit, of the dynamics of an electron coupled to the electromagnetic 
fields, its self-fields included,  basically limits the physically 
sensible choices of initial data to a stationary electron well-separated 
from some localized radiation field that has compact support in space 
disjoint from the fixed support of the particle.
	To have a dynamically interesting scenario, the time-evolved 
support of the initial radiation fields should eventually overlap 
with the support of the electron.
        \section{Gyroscopic LED}
	We study first the subsystem of equations 
obtained by neglecting the world-line equations \refeq{eq:powerEQ}, 
\refeq{eq:NewtonEQ} from the LED with a straight world-line.
	For obvious reasons, we will call this dynamical model the
\emph{gyroscopic} LED.

	We have to solve the Maxwell--Lorentz equations  
\refeq{eq:MLevolvEQfoli}, \refeq{eq:MLdivEQfoli} for the field 
\refeq{eq:EBfoli} together with the gyrograph equations 
\refeq{eq:dualggEQ}, \refeq{eq:torqueVECT} for the bare spin 
\refeq{eq:barespinVECT}. 
	Our strategy is to solve first the Maxwell--Lorentz equations 
in terms of integral representations involving the unknown bare spin 
dynamics.
	Inserting this representation into the gyrograph equation,
we rewrite the latter into a fixed point problem for $\sVb(t)$. 
	We then prove that the fixed point map is a Lipschitz map, 
from which the global well-posedness of the gyroscopic problem follows.
	Subsequently we will show that 
the gyroscopic problem conserves the energy,  angular momentum 
and the canonical spin magnitude, but generally not the linear momentum.
	Energy conservation is coincidental with the
fact that \refeq{eq:powerEQ} is automatically satisfied by a gyroscopic
solution. 
\subsection{Forward integration of the Maxwell-Lorentz equations}
	We recall that in virtue of the 
homogeneous Maxwell--Lorentz equations \refeq{eq:homMLeq}, there 
exists a (non-unique) four-vector field $\qv{A}$ 
 satisfying the  Lorentz gauge $\nabQ\, \cdot \qv{A} = 0$ such that 
$\FQ = \nabQ\wedge \qv{A}$.
	The inhomogeneous Maxwell--Lorentz equation \refeq{eq:inhMLeq} 
then becomes the inhomogeneous wave equation $\wop\qv{A}(\xQ)=4\pi\JQ(\xQ)$.
 	Recalling furthermore the time-plus-space decomposition for the 
current density four-vector, $\JQ(\xQ)=(1,\eulerV(t)\times \xV)\fe(|\xV|)$, 
and introducing the time-plus-space decomposition for the electromagnetic 
potential four-vector as $\AQ(\xQ)=(\phi(\xV,t),\AV(\xV,t))$, the 
equation $\FQ = \nabQ\wedge \qv{A}$ becomes
\begin{equation}
 \GV(\xV,t)=-\nab\phi(\xV,t)- \pdt \AV(\xV,t) + i\nab\times\AV(\xV,t) .
\end{equation}
		The Coulomb potential $\phi$ and vector potential $\AV$
satisfy the inhomogeneous wave equations
\begin{equation}
	\wop \phi(\xV,t) = 4\pi\fe(|\xV|)
\,,\label{eq:waveeqPHI}
\end{equation}
\begin{equation}
	\wop \AV(\xV,t)  = 4\pi\fe(|\xV|)\,\eulerV(t)\times \xV
\, ,\label{eq:waveeqA}
\end{equation}
supplemented (i) by the asymptotic conditions 
$\phi(\xV,t) \sim - e|\xV|^{-1}$ and 
$\AV(\xV,t) \sim \vect{\mu}_0\!\times \xV\abs{\xV}^{-3}$
as $|x|\to\infty$, for all $t\in\Rset$, where $\vect{\mu}_0$
is the particle's magnetic moment at $t = 0$,
\begin{equation}
        \vect{\mu}_0
= 
	\frac{1}{2} \int_{\mathbb{R}^3}
	\xV\times(\eulerV_0\times\xV)\fe(|\xV|)\dd^3x
\, \label{eq:magneticM}
\end{equation}
with $\eulerV_0 = \eulerV(0)$, 
and (ii) by compatible Cauchy data at $t=0$.

	We first integrate the wave equations 
\refeq{eq:waveeqPHI}, \refeq{eq:waveeqA} for the potentials $\phi$ and $\AV$.
	Clearly, \refeq{eq:waveeqPHI} is solved by 
$\phi(\xV,t) = 	\phi_{\text{Coul}}(\xV) + \phi_{\text{wave}}(\xV,t)$,
where 
\begin{equation}
	\phi_{\text{Coul}}(\xV) 
= 
  \int_{\Rset^3}\frac{1}{\abs{\xV-\yV}}\fe\big(\abs{\yV}\big)\,\dd^3y
\label{eq:CoulPOT}
\,\end{equation}
is the static Coulomb potential for $\fe$ 
and $\phi_{\text{wave}}(\xV,t)$ is a solution of the homogeneous scalar
wave equation $\wop \phi_{\text{wave}}(\xV,t) =0$. 
	After at most a  gauge transformation, we may assume that
$\phi_{\text{wave}}\equiv 0$. 
	Next, \refeq{eq:waveeqA} for $t>0$ is solved by
$\AV(\xV,t)=\AV_{\text{source}}(\xV,t) + \AV_{\text{wave}}(\xV,t)$,
where 
\begin{equation} 
\!\!\!
\!\!\!\!\!\!
	\AV_{\text{source}}(\xV,t) 
= 
	\!\!{\int_{\Rset^3}}
\Big(\eulerV_0+\Theta(t-|\xV-\yV|)\big(\eulerV(t-|\xV-\yV|)-\eulerV_0\big)\Big)
	\!\times\!\frac{\yV}{\abs{\xV-\yV}} \fe\big(\abs{\yV}\big)\dd^3y 
\label{eq:AMPEREsolSOURCE}
\end{equation}
solves the inhomogeneous vector wave equation \refeq{eq:waveeqA} 
($\Theta$ is the Heaviside function), and where $\AV_{\text{wave}}(\xV,t)$
solves the homogeneous vector wave equation $\wop \AV_{\text{wave}}(\xV,t) =0$
for initial data $\AV_{\text{wave}}(\xV,0) = \AV^\prime_0(\xV)$
and $\pdt\AV_{\text{wave}}(\xV,0) = - \EV_0^\prime(\xV)$, where
$\AV_0^\prime(\xV) = \AV_0(\xV) - \AV_{\text{source}}(\xV,0)$, with
$\AV_0(\xV)$ the initial magnetic vector potential, and where
$\EV_0^\prime(\xV) = \EV_0(\xV) +\nab\phi_{\text{Coul}}(\xV)$,
with $\EV_0(\xV)$ the initial electric field strength. 
	Thus $\AV_{\text{wave}}$ is given by Kirchhoff's formula
\begin{equation} 
 \AV_{\text{wave}}(\xV,t) 
=
	-\displaystyle{\frac{1}{t}\int_{\partial B_t(\xV)}}
	\EV_0^\prime(\yV) \dd\Omega_{\yV}
+
 \displaystyle{\frac{\partial}{\partial t}
		\Big(\frac{1}{t}\int_{\partial B_t(\xV)}}
\AV_0^\prime(\yV) \dd\Omega_{\yV}\Big)
\, ,\label{eq:KIRCHHOFFrep}
\end{equation}
where $\dd\Omega_{\yV}$ is the uniform surface measure on $\partial
B_t(\xV)$ divided by $4\pi$.
 
        \subsection{Canonical form of the gyrograph equation}
	With the help of the potential representation of $\GV$ 
we now rewrite \refeq{eq:dualggEQ}, \refeq{eq:torqueVECT} into 
the more accessible canonical format. 
	Recalling that 
$\EV(\xV,t) = -\nab\phi_{\text{Coul}}(\xV) - \pdt\AV(\xV,t)$, 
with $\phi_{\text{Coul}}(\xV)$ given in \refeq{eq:CoulPOT}, and
with $\BV(\xV,t) = \nab\times\AV(\xV,t)$, and noticing that 
$\xV\times\nab\phi_{\text{Coul}}(\xV)=\vect{0}$, we find 
\begin{equation}
\begin{array}{rl}
\!\!\!\!\!\!\!\!\!\!\!\!\!\!\!\!
\tV(t)\! &\!\!\!
 = \displaystyle{  \int_{\Rset^3} }
	\xV\times \Big(-\pdt\AV(\xV,t) 
	+ (\eulerV(t)\times\xV)\times\nab\times\AV(\xV,t)\Big)
	{\fe}(|\xV|)\,\dd^3x 
\\ \!\!\!\!\! &\!\!\! =
	-\displaystyle{\ddt  \int_{\Rset^3}}
 	\xV\times \AV(\xV,t) 	{\fe}(|\xV|)\,\dd^3x 
	+ \!\! \int_{\Rset^3} 
	\xV\times \Big( (\eulerV(t)\times\xV)\times\nab\times\AV(\xV,t)\Big)
	{\fe}(|\xV|)\,\dd^3x. 
\end{array}
\label{eq:torqueREWRITEa}
\end{equation}
	The last term in \refeq{eq:torqueREWRITEa} can be rewritten as
\begin{equation}
\!\!\! \!\!\! 
\displaystyle{	\int_{\Rset^3}}
\!\!\! 
\xV\times \big( (\eulerV(t)\times\xV)\times\nab\times\AV(\xV,t)\big)
	{\fe}(|\xV|)\,\dd^3x 
 =
	 \eulerV(t)\times\displaystyle{\int_{\Rset^3} }
\xV \times \AV (\xV,t) \fe(|\xV|)\dd^3x.
\label{eq:torqueREWRITE}
\end{equation}
	To verify \refeq{eq:torqueREWRITE}, first note that
$\xV\times\big((\eulerV(t)\times\xV)\times\nab\times\AV(\xV,t) \big)
= 
 \eulerV(t)\times \xV \big(\xV\cdot\nab\times\AV(\xV,t)\big)$
(for $\xV\cdot(\omegaV\times\xV)={0}$) and pull 
$\eulerV(t)\times$ in front of the integral, next 
use $\nab\times \xV = \vect{0}$ and another standard identity from
vector analysis to rewrite
$\xV\cdot\nab\times \AV =\xV\cdot\nab\times \AV -\AV\cdot\nab\times \xV 
= \nab\cdot (\AV\times \xV)$, then 
integrate by parts, use the identity
$\big(\xV\times\AV(\xV,t)\big)\cdot \nab {\fe}(|\xV|) = 0$
and get
\begin{equation}
\begin{array}{rl}
\displaystyle{\int_{\Rset^3} }
		\xV \big(\xV\cdot\nab\times\AV(\xV,t)\big)
	{\fe}(|\xV|)\,\dd^3x 
&\!\! =
	 \displaystyle{\int_{\Rset^3}}
	\big(\xV\times\AV(\xV,t)\big)\cdot \nab
	\big(\xV{\fe}(|\xV|)\big)\,\dd^3x \\
&\!\! =
	 \displaystyle{\int_{\Rset^3}}
	\xV\times\AV(\xV,t){\fe}(|\xV|)\,\dd^3x,
\end{array}
\label{eq:torqueREWRITEaux}
\end{equation}
as claimed.
	Defining now the electromagnetic field spin vector of the particle by
\begin{equation}
	\sVe(t)
=
	\int_{\Rset^3} \xV \times \AV (\xV,t) \fe(|\xV|)\dd^3x
\, , \label{eq:electromagSPINvect}
\end{equation}
and its canonical spin vector by $\sV =\sVb + \sVe$, and 
finally recalling that $\eulerV\times \sVb = \vect{0}$,
we conclude that \refeq{eq:dualggEQ} can be recast into 
the canonical evolution equation for the spin (in $\labF$),
\begin{equation}
	\ddt {\sV}
=  
	\eulerV\times{\sV}
\, .\label{eq:spinDIFFeq}
\end{equation}
\textbf{Remark:} 
\textit{It follows directly from \refeq{eq:spinDIFFeq} that 
$|\sV|$ is conserved during the evolution.}
        \subsection{The bare spin / angular velocity relation}
	Inserting the explicit integral representation for $\AV(\xV,t)$
into  the canonical equation \refeq{eq:spinDIFFeq}, and recalling 
that $\sVb(t)$ is given in terms of $\eulerV(t)$ by \refeq{eq:barespinVECT}, 
we see that \refeq{eq:spinDIFFeq} becomes a closed, non-autonomous, 
nonlinear first-order vector differential equation for $\eulerV(t)$. 
	However, it is advisable to eliminate $\eulerV(t)$ in favor 
of $\sVb(t)$. 

	We rewrite \refeq{eq:barespinVECT} as
$\sVb(t) = \Iz(|\eulerV(t)|)\cdot\eulerV(t)$, where
\begin{equation}
  \Iz(\abs{\eulerV})
=
  \int_{\Rset^3}
  \frac{|\xV|^2 \mathbf{1} - \xV\otimes\xV}
  {\sqrt{\displaystyle 1-\abs{\eulerV\times\xV}^2}}
  \fm\big(\abs{\xV}\big)  \dd^3x, 
  \label{eq:baremomentofinertia}
\end{equation}
is the inertia tensor of the bare particle. 
	Clearly, $\Iz$ acts as a number on $\eulerV$,
viz. $\Iz\cdot\eulerV =\rotI\eulerV$.
	Performing the angular integrations we are left with
\begin{equation}
	\rotI(|\eulerV|)
=
	2\pi\frac{1}{|\eulerV|^4}\int_0^{|\eulerV|R}
	\fm\left(\frac{\xi}{|\eulerV|}\right)
	\Big(\left(\xi^2+1\right)\Artanh (\xi)-\xi\Big)\xi\dd \xi        
\, .\label{eq::bareINERTmoment}
\end{equation}
	By hypothesis, $0< \rotI(0) < \infty$.  
	This implies that the map $|{\eulerV}|\mapsto{\rotI}(|{\eulerV}|)$
is strictly positive, increasing, and strictly convex 
for $|{\eulerV}|\in[0,1/R)$.
	Depending on the choice for $\fm$, 
the bare spin magnitude $|\sVb|$ may or may not approach a finite limit 
$\sbstar$ as $|\eulerV|R\nearrow 1$.
	In any event, it follows that for ${|\sVb| < \sbstar}$ we can 
invert the map $\eulerV\mapsto\sVb = {\rotI}(|{\eulerV}|)\eulerV$
to get the Euler angular velocity vector $\eulerV$
uniquely in terms of the bare spin vector $\sVb$, viz.
$\eulerV = \WV(\sVb)$, where
\begin{equation}
	\WV(\sVb) 
= 
	\frac{\sVb}{|\sVb|} ({\rotI}\, {\rm id})^{-1}(|\sVb|) 
		\qquad {\rm for}\ \ {|\sVb| < \sbstar}.
\end{equation}
	Note that the map 
$|\sVb|\mapsto (\rotI\, {\rm id})^{-1}(|\sVb|)$ 
is  bounded, strictly increasing, and concave, hence it has 
its steepest slope when $|\sVb|\to 0^+$.
	This slope at $0^+$ is simply the reciprocal value of the slope 
of the map 
$|\omegaV|\mapsto |\omegaV| \rotI(|\omegaV|)$ at $|\omegaV|\to 0^+$, 
viz. slope of $(\rotI\, {\rm id})^{-1} \leq \rotI(0)$ ($< \infty$, 
for $\rotI(0) >0$, by hypothesis). 
	Finally, if $\sbstar<\infty$, we 
extend $\WV$ continuously differentiably to $\Rset^3$ by setting
\begin{equation}
	\WV(\sVb) 
\defeg 
	\frac{1}{R} \frac{\sVb}{|\sVb|} 
		\qquad {\rm for}\ \ {|\sVb| \geq \sbstar}. 
\end{equation}

	\subsection{Bare spin evolution as fixed point problem}
	Substituting $\WV(\sVb)$ for $\eulerV$ in \refeq{eq:spinDIFFeq} 
and integrating \refeq{eq:spinDIFFeq} w.r.t. $t$, 
supplementing the initial datum $\sVb(0)$ (automatically 
compatible with the {subluminality requirement} 
$|\eulerV_0|R \leq 1$), and writing out dependencies on $\AV$ 
explicitly, we arrive at the following integral equation for $\sVb$, 
\begin{equation}
\begin{array}{rl}
	\sVb(t) 
&\!\!\!
= 
	\sVb(0) + 
\displaystyle{\int_{\Rset^3}} 
\xV\times \Big( \AV_0(\xV) - \AV(\xV,t)\Big)\fe(|\xV|)\,\dd^3x 
\\ &
\quad +   \displaystyle{\int_0^t} \WV\big(\sVb(\tilt)\big)\times
   \displaystyle{\int_{\Rset^3}} 
\xV\times\AV(\xV,\tilt)\fe(|\xV|)\,\dd^3x 
	\,\dd \tilt ,
\end{array}
\label{eq:barespinFIXptEQ}
\end{equation}
where $\AV = \AV_{\text{wave}} + \AV_{\text{source}}$ is given 
by the integral representations 
\refeq{eq:KIRCHHOFFrep} and \refeq{eq:AMPEREsolSOURCE},
and where $\eulerV(t) = \WV\big(\sVb(t)\big)$
in \refeq{eq:AMPEREsolSOURCE}, closing the chain.
	Substituting \refeq{eq:AMPEREsolSOURCE} for
$\AV_{\text{source}}$ in \refeq{eq:barespinFIXptEQ} 
and rearranging some integrations gives the explicit
fixed point equation for $\sVb$, 
\begin{equation}
\begin{array}{rl}
	\sVb(t)
= 
&\!\!\!
	\sVb(0) 
	+   \displaystyle{\int_{\Rset^3}} \xV\times
	\big(\AV_{\text{wave}}(\xV,0)
			-\AV_{\text{wave}}(\xV,t)\big)\fe(|\xV|)\,\dd^3x 
\\ &
    -
\displaystyle{\int_0^t}
	\Big(\WV\big(\sVb(\tilt)\big)-\eulerV_0\Big)K(t-\tilt)\dd\tilt
\\ &
	+   \displaystyle{\int_0^t } 
\WV\big(\sVb(\tilt)\big)\times   \displaystyle{\int_{\Rset^3}}
	\xV\times\AV_{\text{wave}}(\xV,\tilt)\fe(|\xV|)\dd^3x\,\dd\tilt 
\\ &
	-\eulerV_0 \times  \displaystyle{\int_0^t}\WV\big(\sVb(\tilt)\big)
	  \displaystyle{\int_\tilt^{2R}}K(t^\prime)
			\dd t^\prime\dd\tilt 
\\ &
	+   \displaystyle{\int_0^t} \WV\big(\sVb(\tilt)\big)\times
	  \displaystyle{\int_0^\tilt}
\WV\big(\sVb(t^\prime)\big)K(\tilt-t^\prime)\dd{t}^\prime\,\dd \tilt ,
\end{array}
\label{eq:sbFIXptEQ}
\end{equation}
where $K$ is the electron's retarded self-interaction kernel,
\begin{equation}
	K(t) 
= 
	\frac{2}{3}
	\int_{\Rset^3}	\int_{\Rset^3}
	\frac{\xV\cdot\yV}{|\xV-\yV|}
	\fe(|\xV|)\fe(|\yV|) \delta(t-|\xV-\yV|) \dd^3x\, \dd^3y.
\label{eq:modulatorFCTgeneral}
\end{equation}
	Notice that $K\in L^\infty(\Rset)$, and that 
$\supp(K)\subseteq [0,2R]$.
	By the $SO(3)$ invariance of $\fe$ we can carry out 
the angular integrations in \refeq{eq:modulatorFCTgeneral}, obtaining a 
double integral,
\begin{equation}
	K(t) 
= 
	\frac{8\pi^2}{3}
	\int_0^R\int_0^R
	\Theta(t-|r-s|)\Theta(r+s-t)
	(r^2 +s^2 - t^2)
	rs \fe(r)\fe(s) \dd r \dd s.
\label{eq:modulatorFCT}
\end{equation}
        \subsection{Lipschitz estimates}
\noindent
\textbf{Lemma 1}: \textit{ The map $\WV:\Rset^3\to\Rset^3$ 
is Lipschitz continuous for the standard Euclidean norm, 
with Lipschitz constant} $1/\rotI(0)$.
\smallskip

\textit{Proof.} 
	The sole action of $\WV$ is to scale
any input vector $\uV$ by the factor $|\WV(\uV)|/|\uV|$, with 
	$|\WV(\uV)| = (\rotI\, {\rm id})^{-1}(|\uV|)$
for $|\uV| < \sbstar$, and $|\WV(\uV)| = 1/R$ for $|\uV|\geq \sbstar$.
	The map $|\uV| \mapsto (\rotI\, {\rm id})^{-1}(|\uV|)$ is
increasing and  concave, vanishing with finite slope $1/\rotI(0)$ 
for $|\uV| \to 0^+$, and saturating  for $|\uV| \to \sbstar$ to
$(\rotI\, {\rm id})^{-1}(|\uV|)\to 1/R$, with vanishing slope.
	Thus, $|\WV(\uV)|/|\uV|$ is monotonic decreasing and bounded 
above by $\lim_{|\uV| \to 0^+}(|\WV(\uV)|/|\uV|) = 1/\rotI(0)$.
	Hence, the map $\rotI(0)\WV$ shrinks any input vector $\uV$ 
by a factor which is the smaller the longer $\uV$ is, but leaving
its direction unchanged.
	It now follows right away that
$|\rotI(0)\WV(\uV_1)- \rotI(0)\WV(\uV_2)| \leq |\uV_1- \uV_2|$
for any two vectors $\uV_1$ and $\uV_2$. 
\hfill{{QED}}
\medskip

\noindent
\textbf{Lemma 2}: \textit{ The two-point map 
$\WV^{\times{2}}:\Rset^3\times\Rset^3\to\Rset^3$ defined by
$(\uV,\vV) \mapsto \WV(\uV)\times\WV(\vV)$ 
is Lipschitz continuous with Lipschitz constant} $(\rotI(0)R)^{-1}$.

\textit{Proof.} 
By a simple identity, followed by the triangle inequality, 
followed by the upper bound $|\WV|\leq 1/R$ and by Lemma 1, we find
\begin{equation}
\begin{array}{rl}
&\!\! 
\displaystyle{
\Big|\WV(\uV_1)\times\WV(\vV_1) 
		-\WV(\uV_2)\times\WV(\vV_2)\Big| }
\\ \\ &\ =
\displaystyle{
	\Big|
	\WV\big(\uV_1\big)\times\Big(\WV\big(\vV_1\big)
				-\WV\big(\vV_2\big)\Big)
	+    
	\Big(\WV\big(\uV_1\big)
	-\WV\big(\uV_2\big)\Big)\times\WV\big(\vV_2\big)\Big|}
\\ \\ &\ \leq
\displaystyle{
	\abs{\WV\big(\uV_1\big)}
	\abs{\WV\big(\vV_1\big) -\WV\big(\vV_2\big)}
	+    
	\abs{\WV\big(\uV_1\big) -\WV\big(\uV_2\big)}
	\abs{\WV\big(\vV_2\big)} }
\\ \\ &\ \leq
\displaystyle{
	R^{-1}
	\Big(
	\abs{\WV\big(\vV_1\big) -\WV\big(\vV_2\big)}
	+    
	\abs{\WV\big(\uV_1\big) -\WV\big(\uV_2\big)}
	\Big)}
\\ \\ &\ \leq
\displaystyle{
	\big(R\,\rotI(0)\big)^{-1}
	\big(
	\abs{\uV_1 -\uV_2}
	+    
	\abs{\vV_1 -\vV_2}
	\big)}\hfill{\mathrm{QED}}
\end{array}\nonumber
\end{equation}

Writing \refeq{eq:sbFIXptEQ} as $\sVb= \FV(\sVb)$ defines
a map $\FV$ in the space $L^1_\lambda(\Rset^+,\Rset^3)$ of $\Rset^3$-valued
functions $\uV$ on $\Rset^+$, equipped with the weighted $L^1$ norm 
$\norml{\uV}=\int_0^\infty \exp(-\lambda t)|\uV(t)|\dd t$, $\lambda >0$.
	Since by assumption the integral of the wave fields over the 
particle support is bounded for all $t$, there exist two 
constants $C_1$ and $C_2$, determined by the initial data alone, such 
that $|\FV(\uV)| < C_1 + C_2t$ for any 
$\uV\in L^1_\lambda(\Rset^+,\Rset^3)$. 
	Hence, $\FV$ maps $L^1_\lambda(\Rset^+,\Rset^3)$ into some ball 
$\{\norml{\uV}\leq C\}\subset L^1_\lambda(\Rset^+,\Rset^3)$, 
where $C$ is determined by the initial data.
	This also implies that $\norml{\sVb}$ is well defined for any
solution of \refeq{eq:sbFIXptEQ}.
\medskip

\noindent
\textbf{Proposition 1}: 
\textit{
The map $\uV \mapsto \FV(\uV)$ is $\norml{\ .\ }$-Lipschitz continuous
with Lipschitz constant}
\begin{equation}
	L 
=
	\frac{1}{\lambda\rotI(0)}
	\left(
	\norm{\sV_{\text{wave}}}_\infty
	+
	\left(1+ \frac{1}{\lambda R}\right)\norm{K}_\infty
	+
	\frac{2}{R}\norm{K}_1
	\right)	 ,
\label{eq:LipschitzL}
\end{equation}
\textit{where}
$\norm{\sV_{\textrm{wave}}}_\infty = \sup_{t}\abs{{\int_{\Rset^3}}
 \xV\times\AV_{\textrm{wave}}(\xV,t)\fe(|\xV|)\dd^3x}<\infty$;
$\norm{K}_\infty = \sup_{t\in [0,2R]}\abs{K(t)}<\infty$, and
$\norm{K}_1 = \int_0^{2R}\abs{K(t)}\dd t<\infty$.
\medskip

\textit{Proof.}
	By definition of $\FV$, 
\begin{equation}
\begin{array}{rl}
&
\!\!\! 
\!\!\! 
\!\!\! 
\FV(\uV)(t)-\FV(\vV)(t)\! = 
    - \displaystyle{\int_0^t}
   \Big(\WV\big(\uV(\tilt)\big)-\WV\big(\vV(\tilt)\big)\Big)K(t-\tilt)\dd\tilt
\\ &\ \ \qquad
	+ \displaystyle{\int_0^t } 
\Big(\WV\big(\uV(\tilt)\big) - \WV\big(\vV(\tilt)\big)\Big)
\!\times\!
   \displaystyle{\int_{\Rset^3}}
	\xV\times\AV_{\text{wave}}(\xV,\tilt)\fe(|\xV|)\dd^3x\dd\tilt 
\\ &\ \ \qquad
    -\eulerV_0 \times  \displaystyle{\int_0^t}
	\Big(\WV\big(\uV(\tilt)\big)-\WV\big(\vV(\tilt)\big)\Big)
	  \displaystyle{\int_\tilt^{2R}}K(t^\prime)
			\dd t^\prime\dd\tilt 
\\ &\ \ \qquad
	+ \displaystyle{\int_0^t} 
	  \displaystyle{\int_0^\tilt}
	\Big(\WV\big(\uV(\tilt)\big)
	\times
	\WV\big(\uV(t^\prime)\big)
-
	\WV\big(\vV(\tilt)\big)
	\times
	\WV\big(\vV(t^\prime)\big)\Big)
		K(\tilt-t^\prime)\dd{t}^\prime\,\dd \tilt .
\end{array}\nonumber
\end{equation}
	Subadditivity of the norm gives
\begin{equation}
\begin{array}{rl}
&\!\!\! \!\!\!\!\!\! \!\!\! 
{\norml{\FV(\uV)-\FV(\vV)}}
 \leq 
\norml{
     \displaystyle{\int_0^t}
   \Big(\WV\big(\uV(\tilt)\big)-\WV\big(\vV(\tilt)\Big)K(t-\tilt)\dd\tilt
	}
\\ & \qquad\qquad + \norml{
	 \displaystyle{\int_0^t } 
	\Big(\WV\big(\uV(\tilt)\big) - \WV\big(\vV(\tilt)\big)\Big)
	\times
	\displaystyle{\int_{\Rset^3}}
	\xV\times\AV_{\text{wave}}(\xV,\tilt)\fe(|\xV|)\dd^3x\dd\tilt 
      }
\\ & \qquad +
\norml{
	\eulerV_0\times \displaystyle{\int_0^t}
\Big(\WV\big(\uV(\tilt)\big)-\WV\big(\vV(\tilt)\big)\Big)
	  \displaystyle{\int_\tilt^{2R}} K(t^\prime)
			\dd t^\prime\dd\tilt 
	}
\\ &
	+ 
\norml{
	\displaystyle{\int_0^t} 
	\displaystyle{\int_0^\tilt}
		\Big(\WV\big(\uV(\tilt)\big)
	\times
		\WV\big(\uV(t^\prime)\big)
-
		\WV\big(\vV(\tilt)\big)
	\times
		\WV\big(\vV(t^\prime)\big)\Big)
		K(\tilt-t^\prime)\dd{t}^\prime\,\dd \tilt 
	}.
\end{array}\nonumber
\end{equation}
	We now estimate one by one the terms on the right-hand side.
	For the first term we find
\begin{equation}
\begin{array}{rl}
&\!\!\! 
\norml{
	\displaystyle{\int_0^t}
	\Big(\WV\big(\uV(\tilt)\big)-\WV\big(\vV(\tilt)\Big)K(t-\tilt)
	\dd\tilt
	}
\\ \leq &
	\displaystyle{\int_0^\infty} e^{-\lambda t}
	\displaystyle{\int_0^t}
	\abs{\WV\big(\uV(\tilt)\big)-\WV\big(\vV(\tilt)\big)}\abs{K(t-\tilt)}
	\dd\tilt \,\dd t
\\ \leq &
	\norm{K}_\infty
	\displaystyle{\int_0^\infty} e^{-\lambda t}
	\displaystyle{\int_0^t}
	\abs{\WV\big(\uV(\tilt)\big)-\WV\big(\vV(\tilt)\big)}\dd\tilt
	\,\dd t
\\ = &
	\norm{K}_\infty
	\lambda^{-1}
	\displaystyle{\int_0^\infty} e^{-\lambda t}
  \abs{\WV\big(\uV(t)\big)-\WV\big(\vV(t)\big)}	\,\dd t
\\ \leq &
	\norm{K}_\infty 
	\big(\lambda\rotI(0)\big)^{-1}
	\displaystyle{\int_0^\infty} e^{-\lambda t}
	  \abs{\uV(t)-\vV(t)}\,\dd t
\\ =    &
	\norm{K}_\infty \big(\lambda\rotI(0)\big)^{-1} \norml{\uV-\vV},
\end{array}\nonumber
\end{equation}
where in the third step we used integration by parts together with
$|\WV|< 1/R$ and with $te^{-\lambda t}=0$ for $t=0$ 
and $t\to \infty$. 
	The last estimate then is Lemma 1.
	Similarly, for the second term we find
\begin{equation}
\begin{array}{rl}
&\!\!\!\!\!\!
 \norml{
	 \displaystyle{\int_0^t } 
	\Big(\WV\big(\uV(\tilt)\big) - \WV\big(\vV(\tilt)\big)\Big)
	\times
	\displaystyle{\int_{\Rset^3}}
	\xV\times\AV_{\text{wave}}(\xV,\tilt)\fe(|\xV|)\dd^3x\dd\tilt 
      }
\\ \leq &
	 \displaystyle{\int_0^\infty} e^{-\lambda t}
	 \displaystyle{\int_0^t } 
	\Big|\WV\big(\uV(\tilt)\big) - \WV\big(\vV(\tilt)\big)\Big|
	\abs{\displaystyle{\int_{\Rset^3}}
	\xV\times\AV_{\text{wave}}(\xV,\tilt)\fe(|\xV|)\dd^3x}\dd\tilt\,
	\dd t
\\ \leq &
	\norm{\sV_{\text{wave}}}_\infty
	 \displaystyle{\int_0^\infty} e^{-\lambda t}
	 \displaystyle{\int_0^t } 
	\Big|\WV\big(\uV(\tilt)\big) - \WV\big(\vV(\tilt)\big)\Big|
	\dd\tilt\,\dd t
\\ = &
	\norm{\sV_{\text{wave}}}_\infty
	\lambda^{-1}
	 \displaystyle{\int_0^\infty} 	 
	e^{-\lambda t}
	\Big|\WV\big(\uV(t)\big) - \WV\big(\vV(t)\big)\Big| \dd t
\\ \leq &
	\norm{\sV_{\text{wave}}}_\infty
	 \big(\lambda\rotI(0)\big)^{-1}
	 \displaystyle{\int_0^\infty} 	 
	e^{-\lambda t}	\big|\uV(t) - \vV(t)\big| \dd t
\\ = &
	\norm{\sV_{\text{wave}}}_\infty
	 \big(\lambda\rotI(0)\big)^{-1}
	\norml{\uV - \vV}.
\end{array}\nonumber
\end{equation}~Proceeding analogously for the third term, we find
\begin{equation}
\begin{array}{rl}
&\!\!\!
\norml{
	\eulerV_0\times \displaystyle{\int_0^t}
\Big(\WV\big(\uV(\tilt)\big)-\WV\big(\vV(\tilt)\big)\Big)
	  \displaystyle{\int_\tilt^{2R}} K(t^\prime)
			\dd t^\prime\dd\tilt 
	}
\\ \leq &
	|\eulerV_0|
	 \displaystyle{\int_0^\infty} e^{-\lambda t}
	 \displaystyle{\int_0^t}
	\Big|\WV\big(\uV(\tilt)\big)-\WV\big(\vV(\tilt)\big)\Big|
	  \displaystyle{\int_\tilt^{2R}}|K(t^\prime)|
			\dd t^\prime\, \dd\tilt \, \dd t
\\ \leq &
	|\eulerV_0|\norm{K}_1 
	 \displaystyle{\int_0^\infty} e^{-\lambda t}
	 \displaystyle{\int_0^t}
	\Big|\WV\big(\uV(\tilt)\big)-\WV\big(\vV(\tilt)\big)\Big|
	\, \dd\tilt \, \dd t
\\ = &
	|\eulerV_0|\norm{K}_1 
	\lambda^{-1}
	 \displaystyle{\int_0^\infty} e^{-\lambda t}
	\abs{\WV\big(\uV(t)\big)-\WV\big(\vV(t)\big)}
	\, \dd t
\\ \leq &
	|\eulerV_0|\norm{K}_1 
	 \big(\lambda\rotI(0)\big)^{-1}
	\displaystyle{\int_0^\infty} e^{-\lambda t}
	  \abs{\uV(t)-\vV(t)}\,\dd t
\\ = &
	|\eulerV_0|\norm{K}_1 \big(\lambda\rotI(0)\big)^{-1}
	  \norml{\uV-\vV}.
\end{array}\nonumber
\end{equation}
	For the fourth term we need Lemma 2, otherwise
we proceed along the same lines to find
\begin{equation}
\begin{array}{rl}
&\!\!\!\!\!\!\!\!\!\!\!\!\!\!\!\!
\norml{
	\displaystyle{\int_0^t} 
	\displaystyle{\int_0^\tilt}
		\Big(\WV\big(\uV(\tilt)\big)
	\times
		\WV\big(\uV(t^\prime)\big)
-
		\WV\big(\vV(\tilt)\big)
	\times
		\WV\big(\vV(t^\prime)\big)\Big)
		K(\tilt-t^\prime)\dd{t}^\prime\,\dd \tilt 
	}
\\ \!\!\!\!\!\!\!\!
\leq &\!\!\!\!
\displaystyle{\int_0^\infty}\!\! e^{-\lambda t}
	\displaystyle{\int_0^t} 
	\displaystyle{\int_0^\tilt}
		\abs{\WV\big(\uV(\tilt)\big)
	\times
		\WV\big(\uV(t^\prime)\big)
-
		\WV\big(\vV(\tilt)\big)
	\times
		\WV\big(\vV(t^\prime)\big)}
		\abs{K(\tilt-t^\prime)}\dd{t}^\prime\,\dd \tilt\,	\dd t
\\  \!\!\!\!\!\!\!\!
 = &\!\!\!\!
	\lambda^{-1}
\displaystyle{\int_0^\infty}\!\! e^{-\lambda t}
	\displaystyle{\int_0^t} 
		\abs{\WV\big(\uV(t)\big)
	\times
		\WV\big(\uV(t^\prime)\big)
-
		\WV\big(\vV(t)\big)
	\times
		\WV\big(\vV(t^\prime)\big)}
		\abs{K(t-t^\prime)}\dd{t}^\prime\, \dd t
\\ \!\!\!\!\!\!\!\!
\leq &\!\!\!\!
	\big(\lambda{R}\rotI(0)\big)^{-1}
\displaystyle{\int_0^\infty}\!\! e^{-\lambda t}
	\displaystyle{\int_0^t} 
	\Big(
	\abs{\uV(t) -\vV(t)}
	+    
	\abs{\uV(t^\prime) -\vV(t^\prime)}
	\Big)
	\abs{K(t-t^\prime)}\dd{t}^\prime\, \dd t
\\  \!\!\!\!\!\!\!\!
  = &\!\!\!\!
	\big(\lambda{R}\rotI(0)\big)^{-1}
	\displaystyle{\int_0^\infty}\!\! e^{-\lambda t}
	\Big(
	\abs{\uV(t) -\vV(t)}
	\displaystyle{\int_0^t} 
	\abs{K(t^\prime)}\dd{t}^\prime
	\!+\!\!    
	\displaystyle{\int_0^t} 
	\abs{\uV(t^\prime) -\vV(t^\prime)}
	\abs{K(t-t^\prime)}\dd{t}^\prime\Big) \dd t
\\ \!\!\!\!\!\!\!\!
\leq &\!\!\!\!
	\big(\lambda{R}\rotI(0)\big)^{-1}
	\Big(\norm{K}_1\norml{\uV -\vV}
	+    
	\lambda^{-1}\norm{K}_\infty \norml{\uV -\vV}\Big).
\end{array}\nonumber
\end{equation}
	Adding all estimates together and finally noting that
$\abs{\eulerV_0}R \leq 1$, we find that
\begin{equation}
\qquad\qquad\qquad\qquad	
	{\norml{\FV(\uV)-\FV(\vV)}}
 \leq 
	L {\norml{\uV-\vV}}
\nonumber
\end{equation}
with $L$ given in \refeq{eq:LipschitzL}.\hfill{QED}
        \subsection{Global well-posedness}
	The existence of a unique $\norml{\ .\ }$-strong forward solution 
$t\mapsto \sVb(t)$, $t\geq 0$, of \refeq{eq:spinDIFFeq} now 
follows right away from the $\norml{\ .\ }$-Lipschitz continuity 
of $\FV$ and the fact that $\FV$ maps $L^1_\lambda(\Rset^+,\Rset^3)$
into some ball $\norml{\ .\ }\leq C$, with $C$ determined by the 
initial data.
	Moreover, we can exchange $t\to -t$ and the conclusions
holds for the backward  evolution as well. 
	Furthermore, for any permissible incoming data (not necessarily
scattering data) $\AV_{\text{wave}}(\xV,0)$ we can find a $\lambda_*$ 
such that $L<1$ for all $\lambda > \lambda_*$.
	We summarize these findings in the following theorem.

\noindent
\textbf{Theorem 1}: 
\textit{
	There exists a unique $\norml{\ .\ }$-strong solution }
$t\mapsto \sVb(t)$ 
\textit{of \refeq{eq:spinDIFFeq} globally in $t\in \Rset$.
	Furthermore, for all $\lambda >\lambda_*$ the map
$\FV$ is a $\norml{\ .\ }$-contraction mapping, and in these
norms the simple iteration }
\begin{equation}
	\sVb^{(n+1)} 
= 
	\FV\Big(\sVb^{(n)}\Big),
\end{equation}
\textit{starting with initial datum} 
$\sVb^{(0)} \equiv \sVb(0)$, 
\textit{converges $\norml{\ .\ }$-strongly to the solution} $t\mapsto \sVb(t)$.
\newpage

Global well-posedness in $\norml{\ .\ }$ can now be bootstrapped
to higher regularity for $\sVb(t)$, e.g. $C^1$ regularity if
the Cauchy data for $\AV_{\text{wave}}$ and the densities $\fm$ and $\fe$
are of class $C_0^1(\Lambda)$, where $\Lambda\subset\Rset^3$  is compact. 
	The regularity of $\AV(\xV,t)$ follows accordingly.
	Unfortunately, a detailed discussion of higher regularity 
is beyond the scope of this letter and has to be deferred to some later work.
	However, note that analyticity of $t\mapsto\sVb(t)$ cannot hold, first
because of the compactly supported $\AV_{\text{wave}}$, $\fm$ and $\fe$,  
and furthermore, because it takes only a finite amount of energy to spin up 
the particle so that its equatorial velocity reaches the speed of light
when $\fm(\abs{\ .\ })\in L_+^\infty$ (with compact support in 
$B_R\subset\Rset^3$); for $\Mrot(t) = \rotM(|\eulerV(t)|)$, 
where (by the $SO(3)$ invariance of $\fm$)
\begin{equation}
	\rotM(|\eulerV|)
=
	4\pi\frac{1}{|\eulerV|^3}\int_0^{|\eulerV|R}
	\fm\left(\frac{\xi}{|\eulerV|}\right)
	\Artanh (\xi)\xi \dd \xi
, \end{equation}
and we see that $|\rotM(|\eulerV|)| <C$ as $|\eulerV| \nearrow 1/R$
whenever $\fm \in L^\infty_+$.
	\subsection{Conservation laws}
\noindent
\textbf{Proposition 2}:
{\it	The following quantities are conserved during the evolution:}

\noindent
\begin{equation}
-e = \int_{\mathbb R^3} \rho \, \dd^3x  \qquad (charge),
\label{eq:totalcharge}
\end{equation}
\begin{equation}
	W 
= 
	\frac{1}{8\pi}\int_{\mathbb R^3}\bigl(|\EV|^2 +|\BV|^2\bigr)\,\dd^3x 
	+ \rotM(|\eulerV|) \qquad (energy), 
\label{eq:totalenergy}
\end{equation}
\begin{equation}
	\LV
 = 
	\frac{1}{4\pi} \int_{\mathbb R^3} \xV\times(\EV\times\BV) \,\dd^3x 
	+ \sVb
\qquad (angular\ momentum),
\label{eq:totaldrehimpuls}
\end{equation}
\begin{equation}
	\sigma
=  
	\abs{\sVb + \sVe}\qquad (canonical\ spin\ magnitude).
\label{eq:totalSPINvectMOD}
\end{equation}

\textit{Proof.}
	We basically follow~\cite{kiessling} where the 
conservation laws for the semi-relativistic theory are discussed.

	As for charge conservation, by way of
construction~\cite{nodvik}, LED honors the continuity equation 
\begin{equation}
	\partial_t \rho(\vect{x}, t) 
	+ \nab\cdot \jV(\vect{x},t) 
=
	0,
\label{eq:continuityLAW}
\end{equation}
where $\rho$ is the electric charge density and $\jV$ the vector of the
electric current density, and this fact does not change by simply imposing 
the condition that the world-line be straight.
	Indeed, one directly verifies that for our 
$\jV(\xV,t) = \fe(|\xV|)\,\eulerV(t)\times\xV$ we have  
$\nab\cdot\big(\eulerV(t)\times\xV \fe(|\xV|)\big)={0}$, and of course
$\rho(\vect{x},t) = \fe(|\xV|)$ independent of $t$, i.e.
$\pdt\rho(\xV,t)=0$.
	Hence, charge is conserved.

	As for the energy conservation, taking the time derivative of 
the field energy gives us~\cite{jacksonBOOK}
\begin{equation}
	\frac{\dd}{\dd t}\Big(\frac{1}{8\pi}
	\int_{\Rset^3}
	\bigl(|\EV(\vect{x},t)|^2 +|\BV(\vect{x},t)|^2\bigr)\,\dd^3x \Big)
= 
	- \int_{\Rset^3}\EV(\vect{x},t) \cdot\jV(\vect{x},t) \,\dd^3x, 
\end{equation}
here with $\jV(\xV,t) =  \fe(|\xV|)\,\eulerV(t)\times\xV$.  
	On the other hand, by direct calculation with
\refeq{eq:gyroMASSeval} and \refeq{eq:barespinVECT} one readily
verifies that
\begin{equation}
	\ddt\rotM(|\eulerV|) 
%
%
= 
	\eulerV\cdot \ddt\sVb.
\label{eq:energyCONSERVEDa}
\end{equation}
	Next, taking the Euclidean inner product with $\eulerV$ 
on both sides of the canonical evolution equation for the total
spin, \refeq{eq:spinDIFFeq}, we see that
\begin{equation}
	\eulerV\cdot\ddt\sVb =	-\eulerV\cdot\ddt\sVe .
\label{eq:energyCONSERVEDb}
\end{equation}
	Recalling now the definition of the electromagnetic field spin,
\refeq{eq:electromagSPINvect}, then using the cyclicity of
$\eulerV\cdot(\xV\times\pdt\AV)$, noting next 
that $-\pdt\AV =\EV+ \nab\phi_{\text{Coul}}$ and that
$(\eulerV\times\xV)\cdot\nab\phi_{\text{Coul}}(|\xV|)=0$, 
and at last recalling that $\fe(|\xV|)\,\eulerV(t)\times\xV =\jV(\xV,t)$
we find
\begin{equation}
\begin{array}{rl}
\!\!\!\!\!\!	
	-\eulerV(t)\cdot\displaystyle{\ddt}\sVe(t)
&\!\! = 
	-\eulerV(t)\cdot
	\displaystyle{
	\int_{\Rset^3}}\xV\times\pdt\AV(\xV,t)\fe(|\xV|)\dd^3x
\\ &\!\! = 
	\displaystyle{
\int_{\Rset^3}}(\eulerV(t)\times \xV)\cdot\EV (\xV,t)\fe(|\xV|)\dd^3x
= 
 \displaystyle{\int_{\Rset^3}}\EV(\vect{x},t) \cdot\jV(\vect{x},t) \,\dd^3x. 
\end{array}
\label{eq:energyCONSERVEDc}
\end{equation}
	Hence, energy conservation is proved.

	As for the angular momentum conservation, taking the time 
derivative of the field angular momentum gives the well-known 
formula~\cite{jacksonBOOK}
\begin{equation}
\!\!\!\!\!\!\!\!\!\!\!
	\frac{\dd}{\dd t}\Big(\frac{1}{4\pi}
	\int_{\Rset^3}\!\!
\xV\times\bigl(\EV(\vect{x},t)\times\BV(\vect{x},t)\bigr)\,\dd^3x \Big)
\! = \!
-\! \int_{\Rset^3}\!\!
\xV\times\Big(\rho(\vect{x},t)\EV(\vect{x},t) +
			\jV(\vect{x},t)\times\BV(\vect{x},t)\Big)\,\dd^3x.
\end{equation}
	Inserting our expressions $\rho(\xV,t) =  \fe(|\xV|)$ and   
$\jV(\xV,t) =  \fe(|\xV|)\,\eulerV(t)\times\xV$, we see that
\begin{equation}
\int_{\Rset^3}\xV\times\Big(\rho(\vect{x},t) \EV(\vect{x},t) 
 + \jV(\vect{x},t) \times\BV(\vect{x},t) \Big)\,\dd^3x  
= 
	\ddt{\sVb}(t),
\end{equation} 
and conservation of angular momentum  is proven.

	Finally, we already remarked that \refeq{eq:spinDIFFeq} 
implies at once that $|\sV|$ is conserved.\hfill{QED}
\smallskip

	The proof that the total energy is conserved has the following
spin-off.

\noindent
\textbf{Corollary 1}:
\textit{ 
	The constraint equation \refeq{eq:powerEQ} is automatically 
satisfied by any solution of gyroscopic LED.
}

	As for the total linear momentum,	
\begin{equation}
	\PV 
= 
	\frac{1}{4\pi} \int_{\Rset^3} \EV\times\BV \, \dd^3x 
	+
	\mathbf{N}_{\text{e}}\cdot\eulerV
\, , \label{eq:totalimpuls}
\end{equation}
we remark that conservation of \refeq{eq:totalimpuls} is equivalent
to \refeq{eq:NewtonEQ}.
	However, our assumption of a straight particle world-line
is generally \emph{not} compatible with  \refeq{eq:NewtonEQ},
unless special symmetries prevail. 
	An example is discussed in the next section.
	\section{Rotation-reflection symmetric scattering}
	Our Theorem 1 reduces the global existence and uniqueness 
problem for proper LED with a straight particle world-line 
to finding the class of non-stationary initial conditions for which 
momentum conservation holds with a non-moving particle. 
	Such a class of initial conditions is given by  the 
rotation-reflection symmetric field decorations of spacetime, 
with the particle's axis of rotation necessarily 
identical to the axis of symmetry $\vect{a}$, linearly superimposed 
on which is a compactly supported, non-symmetric electromagnetic 
radiation field that is never going to interact with the particle. 
	Since a non-interacting radiation field is evidently 
rather uninteresting,  we confine our discussion to
the rotation-reflection symmetric evolutions.

	More precisely, let $(\zeta, \theta,z)$ denote cylindrical 
coordinates of $\xV$, with origin in the 
particle center, axis unit vector $\vect{a}$, $z = \xV\cdot\aV$,
$\theta$ the polar angle of $\xV$ about $\aV$, and 
$\zeta = |\xV - z\aV|$.
	The axis $\vect{a}$ is fixed during the evolution, 
and $\eulerV =\euler \vect{a}$, so that $\euler$ is (assumed, and
below verified to be) the only remaining dynamical degree of freedom 
of the  particle. 
	Aside from the non-dynamical and spherically symmetric 
Coulomb field \refeq{eq:CoulPOT}, the remaining electromagnetic 
field is now determined by a vector potential of
the form $\AV(\xV,t) = \psi(\zeta,z,t)\nab\theta$, satisfying
the reflection symmetry $\psi(\zeta,z,t)=\psi(\zeta,-z,t)$,
and obviously rotation invariant.
	The inhomogeneous wave  equation for $\AV$,
\refeq{eq:waveeqA}, reduces to the inhomogeneous, 
scalar, generalized wave equation 
\begin{equation}
	\big( \partial_{tt} - \partial_{\zeta\zeta} 
+ \zeta^{-1}\partial_\zeta - \partial_{zz}\big)\psi(\zeta,z,t) 
=
	4\pi \omega(t)\zeta^2 \fe\big(\sqrt{\zeta^2+z^2}\big),
\label{eq:waveeqPSI}
\end{equation}
with accordingly simplified scalar solution formulas for $\psi$. 
	An elementary calculation with 
$\EV = -\nab\phi_{\textrm{Coul}}-\pdt\psi\nab\theta$
and 
$\BV = \nab\psi\times\nab\theta$ then shows that the torque 
$\int \xV\times(\EV\times\BV)\fe\dd^3x\propto\vect{a}$, 
establishing the consistency at the level of the gyroscopic problem, indeed. 
        \subsection{Momentum balance}
	We already saw that the time component \refeq{eq:powerEQ}
of the covariant world-line equation is automatically satisfied, 
see section 5. 
	We now show that for rotation-reflection symmetric solutions 
to the gyroscopic problem the space-part of the world-line 
constraint equation \refeq{eq:NewtonEQ} is satisfied, too.
	Since the fulfillment of \refeq{eq:NewtonEQ} is equivalent to
the conservation of linear momentum \refeq{eq:totalimpuls}, it suffices
to show that \refeq{eq:totalimpuls} is a constant vector for all time.

	By direct computation with 
$\EV = -\nab\phi_{\textrm{Coul}}-\pdt\psi\nab\theta$
and 
$\BV = \nab\psi\times\nab\theta$
one verifies that 
\begin{equation}
	\int_{\Rset^3}\EV(\xV,t)\times \BV(\xV,t) \,\fe(|\xV|) \dd^3x 
=
  -\int_{\Rset^3}\zeta^{-2} \pdt\psi(\zeta,z,t)\nab\psi(\zeta,z,t)\dd^3x
=
	\vect{0}
\end{equation}
for our rotation-reflection symmetric fields. 
	As for the spin-orbit coupling term, another direct
calculation yields that rotation-reflection symmetry implies
\begin{equation}
	\mathbf{N}_{\text{e}}(t)\cdot\eulerV (t)
= 
-\omega(t)\ddt\int_{\Rset^3}\xV\, \psi(\zeta,z,t) \,\fe(|\xV|)  \dd^3x
=
	\vect{0},
\end{equation}
and the satisfaction of the world-line constraint equation
\refeq{eq:NewtonEQ} follows.

        \subsection{Exponential convergence to the soliton state}
	In~\cite{appelkiesslingAOP} we  proved that the conservation of 
$\sigma=|\sVb+\sVe|$ together with  the invertibility of the map 
$\eulerV\mapsto \sV$ in stationary situations implies that any 
scattering process connects two boosted stationary particle states 
with identical values for the renormalized mass and the magnitudes 
of spin  and magnetic moment.
	In short:  \emph{the Lorentz electron scatters like a 
soliton}. 
	We now complement this result by proving that rotation-reflection
symmetric scattering does occur, and that  the soliton 
state is approached exponentially fast.
	For our proof we need to assume that the ratio of
electrostatic to bare rest mass is small.

\noindent
\textbf{Proposition 3}:
\textit{
	Assume that the electromagnetic potential data are rotation-reflection
symmetric in the sense explained above, and of class $C^1$.
	Assume furthermore that} $\psi_{\textrm{wave}}(\zeta,z,0)$
\textit{has compact support a finite distance away from} $\supp(\fe)$. 
\textit{Finally, assume that}
\begin{equation}
	{\norm{K}_1}< {\rotI(0)}.
\label{eq:smallnessCONDITION}
\end{equation}
\textit{Then,  as $t\to\infty$, the bare spin} $\sVb(t)$ 
\textit{converges exponentially fast to a stationary vector,} 
$\sVb(t)\to\sVb^\infty$, 
\textit{and}
$\sVb^\infty=\sVb(0)$.

\textit{Proof.}
	Clearly, since $\eulerV\propto\vect{a}$ for all $t$, 
all terms $\eulerV_0\times\WV(\sVb)$ and
$\WV\big(\sVb(t)\big)\times\WV\big(\sVb(\tilt)\big)$ vanish.
	Also, by direct calculation one verifies that
${\int_{\Rset^3}}
\xV\times\AV_{\text{wave}}(\xV,t)\fe(|\xV|)\,\dd^3x\propto
\vect{a}$ for all $t$, so that its cross product with $\WV$ vanishes
as well for all $t$.
	Furthermore, by hypothesis, the initial wave data don't
overlap with the support of the particle, hence 
${\int_{\Rset^3}} \xV\times
	\AV_{\text{wave}}(\xV,0)\fe(|\xV|)\,\dd^3x =\vect{0}$.
	Finally, by the wave propagation, there exists a $T \geq 2R$ 
such that 
$\supp\big(\AV_{\text{wave}}(\xV,t)\big)\cap\supp\big(\fe(|\xV|)\big)
=\emptyset$ for all $t>T$.
	Then, for $t> T$, we have 
\begin{equation}
	\sVb(t) 
	+ \displaystyle{\int_{t-2R}^t}
	\WV\big(\sVb(\tilt)\big)K(t-\tilt)\dd\tilt
 = 
	\sV(0),
\qquad \textrm{for}\quad t>T
\label{eq:sbFIXptEQtLARGE}
\end{equation}
where $	\sV(0) =\sVb(0)	+\,  \kappa\eulerV_0$, with
$\kappa\defeg {\int_0^{2R}} K(t)\dd{t}$.
	Notice that \refeq{eq:sbFIXptEQtLARGE} is effectively
a scalar equation because all vectors are $\propto\vect{a}$.
	We now define $\sVb^\infty$ as the -- unique -- solution of
\begin{equation}
	\sVb^\infty 
	+ 
	\kappa	\WV\big(\sVb^\infty\big)
 = 
	\sV(0).
\label{eq:sbFIXptEQstar}
\end{equation}
	Clearly, since $\sV(0)=	\sVb(0)	+\, \kappa \eulerV_0$ and
$\WV\big(\sVb(0)\big) =  \eulerV_0$, 
\refeq{eq:sbFIXptEQstar} is solved by $\sVb^\infty = \sVb(0)$, 
and by uniqueness this is the only solution.
	We next rewrite \refeq{eq:sbFIXptEQtLARGE} as
\begin{equation}
	\sVb(t) - \sVb^\infty  
 = 
	- \displaystyle{\int_{t-2R}^t}\Big(
	\WV\big(\sVb(\tilt)\big)- \WV\big(\sVb^\infty\big)
		\Big)K(t-\tilt)\dd\tilt    
\quad \textrm{for}\quad t>T
\label{eq:sbFIXptEQtLARGErev}
\end{equation}
and estimate
\begin{equation}
\begin{array}{rl}
	\abs{\sVb(t) - \sVb^\infty} 
&
\!\leq
	\displaystyle{\int_{t-2R}^t}
	\abs{\WV\big(\sVb(\tilt)\big)- \WV\big(\sVb^\infty\big)}
	\abs{K(t-\tilt)}\dd\tilt
\\ &\! \leq 
	\big(\rotI(0)\big)^{-1}\displaystyle{\int_{t-2R}^t}
	\abs{\sVb(\tilt)- \sVb^\infty}\abs{K(t-\tilt)}\dd\tilt
\\ &\! \leq 
	\norm{K}_1\big(\rotI(0)\big)^{-1}
	\max_{\tilt\in [t-2R,t]}\abs{\sVb(\tilt)- \sVb^\infty}
\end{array}
\label{eq:sbFIXptEQtLARGEmod}
\end{equation}
where we used the Lipschitz continuity of $\WV$ (Lemma 1)
and the continuity of $t\mapsto\sVb(t)$.
	Now assume that $t\in [n2R, (n+1)2R]$, with $n$ big enough so
that $n2R > T$. 
	By \refeq{eq:sbFIXptEQtLARGEmod} and the inclusion
$[t-2R,t]\subset[(n-1)2R,(n+1)2R]$ we have that
\begin{equation}
	\max_{t\in [n2R,(n+1)2R]}\abs{\sVb(t) - \sVb^\infty} 
\leq 
	\norm{K}_1\big(\rotI(0)\big)^{-1}
	\max_{t\in [(n-1)2R,(n+1)2R]}
	\abs{\sVb(t)- \sVb^\infty}
\label{eq:sbFIXptDEVIATION}
\end{equation}
	By our smallness condition \refeq{eq:smallnessCONDITION}
we conclude that
$\max_{t\in [(n-1)2R,(n+1)2R]}\abs{\sVb(t)- \sVb^\infty}$ cannot be 
attained in $[n2R,(n+1)2R]$, hence it is attained in $[(n-1)2R,n2R)$.
	By induction from one interval of length
$2R$ to the next one we now get 
$|\sVb(nT) - \sVb^\infty| \leq C \exp(-n\Gamma)$, i.e.
exponential convergence with rate
$\Gamma = \ln\big(\rotI(0)/\norm{K}_1\big)$. \hfill{QED}

	The  exponentially fast convergence $\sVb(t)\to\sVb^\infty$ implies
for all rotation-reflection symmetric initial conditions of the 
type discussed that the field-particle system in fact converges 
exponentially fast on families of nested compact sets  to a 
stationary particle-field bound state, the soliton state,
while a departing field 
of electromagnetic radiation escapes to spatial infinity.   
	Put differently, our class of rotation-reflection states consists 
of \emph{scattering states}, with the exception of the stationary
bound state itself.
	For late times the evolution of the electromagnetic field thus
satisfies the scattering formulas ($^*$ means complex conjugate)
\begin{equation}
	\GV(\xV,t) 
\quad{\stackrel{t \to +\infty}{\longrightarrow }}\quad
	\GV_{\text{sol}}^{\text{out}}(\xV) 
	+ e^{-it\nab\times}\ \GV^{\text{out}}_{\text{rad}}(\xV) 
\, ,
\end{equation}
and
\begin{equation}
	\GV^*(\xV,t) 
\quad{\stackrel{ t \to -\infty}{\longrightarrow }}\quad
	\GV_{\text{sol}}^{\text{in}}{^*}(\xV) 
	+ e^{it\nab\times}\ \GV_{\text{rad}}^{\text{in}}{^*}(\xV) 
\, ,
\end{equation}
where the soliton fields $\GV^{\text{in}}_{\text{sol}}$ 
and $\GV^{\text{out}}_{\text{sol}}$ coincide in this 
rotation-reflection symmetric setting.

	\section{Open problems}
	It is instructive to have some explicit numbers.
	As in~\cite{appelkiesslingAOP}, consider the example
where $\fe$ and $\fm$ are given by the uniform surface measure 
on a sphere of radius $R$, i.e.
$\fe(\abs{\xV})= {-e}({4\pi R^2})^{-1}\,\delta(\abs{\xV}-R)$,
and 
$\fm(\abs{\xV}) ={\mz}({4\pi R^2})^{-1}\,\delta(\abs{\xV}-R)$,
with $\mz$ the strictly positive \emph{bare rest mass} of the 
Lorentz electron.
	This gives $\rotI(0)=(2/3)\mz{R^2}$, and 
\begin{equation}
	K(t) 
= 
    e^2\frac{1}{3}\Big(1 -\frac{1}{2}\frac{t^2}{R^2}\Big)\Theta(t)\Theta(2R-t)
\, ,\label{eq:modulatorFCTsphere}
\end{equation}
so that $\norm{K}_1 = e^2 R 2(2{\sqrt{2}-1})/9$.
	Our smallness condition  $\norm{K}_1 < \rotI(0)$ then becomes 
\begin{equation}
	\frac{e^2}{\mz R} < \frac{3}{2\sqrt{2}-1}.
\label{eq:massRATIOsmall}
\end{equation}
	Roughly speaking, the particle's electrostatic Coulomb energy
must be less than the bare rest mass. 
	(This conclusion holds with minor numerical differences
also when $\fm$ is uniform volume measure in $B_R$.)
	The interesting question now is whether deviations from the 
soliton state decay exponentially fast also when the 
smallness condition \refeq{eq:massRATIOsmall} is violated, especially
since one is interested in a renormalization flow limit $\mz\to 0^+$ 
where $R\to 1.5\RC$ (with $\RC$ the electron's Compton 
length)~\cite{appelkiesslingAOP}. 
	Conceivably some  long-lived \emph{resonances} may emerge and
render a more complicated picture. 
	Nonlinear resonances have been studied rigorously in the
simpler semi-relativistic model of a particle interacting with 
a scalar wave field~\cite{kunze}; see also~\cite{sofferweinstein}
for certain nonlinear wave equations.
	A corresponding study for gyroscopic LED is in its infancy.

	For general non-rotation-reflection symmetric initial data
we proved global existence and uniqueness of gyroscopic solutions 
(which typically do \emph{not} satisfy the world-line equations of LED), 
but we do not yet know that on families of nested compact sets the 
field-particle system converges to a stationary state.
	All we can show is that $\sVb(t)$ converges to some $\sVb^\infty$ 
as $t\to\infty$ whenever  the  iterated integral
$\int_0^t \int_0^\tilt 
\WV\big(\sVb(\tilt)\big)\times \WV\big(\sVb(t^\prime)\big)
	K(\tilt-t^\prime)\dd{t}^\prime\,\dd \tilt$
has a limit in $\Rset^3$ as $t\to\infty$, but we have nothing
to say about exponentially fast convergence, then.
	In case of a scattering scenario, i.e.
with convergence to a soliton, the fields 
$\GV_{\text{sol}}^{\text{in}}$ and $\GV_{\text{sol}}^{\text{out}}$ 
are generally \emph{not identical};
however, they differ by at most a space rotation 
as a consequence of the soliton dynamics.
	The explicit characterization of the 
scattering operator from the ``in'' states to the ``out'' states
has yet to be worked out.

	Eventually we would like to be able to establish control over the 
problem of \emph{many-body scattering}. 
	While well developed in quantum 
theory~\cite{duerretal, reedsimonBOOKiii, sigalsofferA, sigalsofferB}, 
very little is known rigorously for truly relativistic LED. 
	Interestingly enough, the solution to this problem requires the 
construction of a self-consistent nontrivial foliation of space-time, 
injecting a technical element from general relativity into the analysis.
\smallskip

\noindent
\textbf{ACKNOWLEDGMENT:} 
The authors are grateful to Herbert Spohn and Avraham Soffer for
their enthusiastic support and interest in this work.
Thanks go also to Stefan Teufel for pointing out the benefits
of exponentially weighted norms in the Lipschitz estimates. 
Last not least, we thank Markus Kunze for his comments on the 
manuscript.

\end{document}